\documentclass[12pt]{article}
\usepackage[bookmarks,bookmarksnumbered]{hyperref}
\usepackage{amsmath,XoohmE}
\usepackage{graphicx,color}
\usepackage{booktabs,multirow}

\def\ra{{\rm a}}
\def\A{@}
\def\BA{\boldsymbol{A}}
\def\cA{\mathcal{A}}
\def\sA{\mathscr{A}}
\def\Db{\relax\leavevmode\hbox{$D$\kern-.6em
               \vrule height1.9ex depth-1.8ex width5pt}\kern2pt}
\def\bL{\boldsymbol{\L}}
\def\bl{{\mkern3mu\vrule height1ex depth-.9ex width4pt
          \mkern-9.25mu\lambda}}
\def\sL{\mathscr{L}}
\def\mm{{=}} 
\def\pp{{=\mkern-10mu|\mkern3mu}}
\def\Qb{\relax\leavevmode\hbox{$Q$\kern-.5em
               \vrule height1.9ex depth-1.8ex width5pt}\kern1pt}
\def\bQ{\boldsymbol{\Q}}

\def\vC#1{\vcenter{\hbox{\hss#1\hss}}}
\definecolor{gray}{rgb}{.7,.7,.7}
\def\cb#1#2{\colorbox{#1}{\color{#1}\fbox{\color{black}#2}}}
\def\cB#1{\hbox to0pt{\hss\color{gray}\fbox{\cb{white}{#1}}\hss}}
\def\CB#1{\hbox to0pt{\hss\color{gray}\fbox{\cb{yellow}{#1}}\hss}}

\def\BX{\boldsymbol{X}}
\def\cX{\mathcal{X}}
\def\sX{\mathscr{X}}
\def\BY{\boldsymbol{Y}}
\def\cY{\mathcal{Y}}
\def\sY{\mathscr{Y}}

\definecolor{Green}  {rgb}{0.00,0.75,0.00} 
\definecolor{Red}    {rgb}{1.00,0.00,0.00} 
\definecolor{Blue}   {rgb}{0.00,0.00,1.00} 
\definecolor{Orange} {rgb}{1.00,0.56,0.00} 
\definecolor{Purple} {rgb}{0.50,0.00,0.50} 
\definecolor{Gold}   {rgb}{1.00,0.75,0.25} 
\definecolor{Magenta}{rgb}{1.00,0.00,1.00} 
\definecolor{Turque} {rgb}{0.00,0.88,0.88} 
\definecolor{Seaweed}{rgb}{0.00,0.25,0.00} 
\definecolor{Brown}  {rgb}{0.50,0.13,0.00} 
\definecolor{Cobalt} {rgb}{0.00,0.00,0.50} 
\definecolor{Sage}   {rgb}{0.00,0.50,0.38} 
\definecolor{grey1}  {rgb}{0.20,0.20,0.20} 
\definecolor{grey2}  {rgb}{0.40,0.40,0.40} 
\definecolor{grey3}  {rgb}{0.60,0.60,0.60} 
\definecolor{grey4}  {rgb}{0.80,0.80,0.80} 
\definecolor{grey5}  {rgb}{0.90,0.90,0.90} 
\def\C#1#2{{\ifcase#1\or
             \color{Red}\or\color{Green}\or\color{Blue}\or
              \color{Orange}\or\color{Purple}\or\color{Gold}\or
             \color{Magenta}\or\color{Turque}\or\color{Seaweed}\or
               \color{Brown}\or\color{Cobalt}\or\color{Sage}\or
                 \color{grey1}\or\color{grey2}\or\color{grey3}\or
                 \color{grey4}\else\color{grey5}\fi#2}}

 \SfTitles
 \allowdisplaybreaks
\begin{document}

  \hfill\today\\[0pt]
 \begin{center}
{\LARGE\sf\bfseries\boldmath
  Unidexterously Constrained Worldsheet Superfields
 }\\*[8pt]
{\sf\bfseries T.\,H\"{u}bsch
}\\*[-2pt]
{\small\it
      Department of Physics \&\ Astronomy,\\[-4pt]
      Howard University, Washington, DC 20059\\[-5pt]
{\tt  thubsch@howard.edu}
 }\\[4pt]
{\small\sf\bfseries ABSTRACT}\\[5pt]
\parbox{126mm}{\addtolength{\baselineskip}{-3pt}\parindent=2pc\noindent
Super-constrained superfields have provided for most of the best-known and oft-used representations of supersymmetry. The abelian nature of the Lorentz symmetry on the worldsheet turns out to continue permitting the discovery of new representations, long after the discovery of twisted chiral superfields. Just as the latter were effectively used in Lagrangian studies of mirror symmetry, it would seem hopeful that the new representations discussed herein may find their own niche too.
}
\end{center}
\begin{flushright}\small\sl
If you do not expect the unexpected, you will not find it.\\[-1mm]
---~Aristotle
\end{flushright}

\section{Introduction, Results and Summary}
 \label{s:0}
Supersymmetry has been studied for almost four decades in the physics literature, and about as long using the superspace formalism\cite{rSSSS1,r1001,rPW,rWB,rBK}. Nevertheless, a complete classification of off-shell representations is still in progress, even just on the worldline; see Refs.\cite{r6-1,r6-3,r6-3.2} and\cite{rFKS,rFS1,rFS2,rKRT,rBKMO,rBG,rBKLS}. These results indicate that there is an overabundance of even the restricted class of so-called {\em\/adinkraic\/} supermultiplets, wherein each supercharge transforms each component field into precisely one other component field (or its derivative), not a linear combination of those. The plentitude of these findings\cite{r6-3,r6-3.2} dwarfs the efforts from earlier literature\cite{r1001,rPW,rWB,rHKLR,rBK,rHSS,rGSS}.

In 1+1-dimensional spacetime|worldsheet|models, it is possible to define superfields that are only {\em\/partly on-shell\/} (on {\em\/half-shell\/}\cite{rHP1}), such as:
\begin{equation}
 \textbf{lefton}:\quad D_-\bL=0=\Db_-\bL,\quad\To\quad \vd_\mm\bL=0.
 \label{e:L}
\end{equation}
Such superfields are off-shell on the left-handed light-cone worldline (0-brane) embedded within the worldsheet, but are not off-shell in the usual sense of $1{+}1$-dimensional worldsheet field theory, and were therefore said to be {\em\/on half-shell\/}\cite{rHP1}. This unique feature of worldsheet field theory of course further complicates any classification of supermultiplets and their superfield representation, but also provides opportunities not possible in other dimensions.

It is our present purpose to demonstrate that unidexterous Lagrange multiplier superfields may be used in two complementary roles, to define:
 ({\small\bf1})~certain proper off-shell superfields which conventional super-differential constraining cannot (see Section~\ref{s:Strict}), and
 ({\small\bf2})~partially on-shell superfields, with an unequal number of off-shell bosonic and fermionic component fields (see Section~\ref{s:Limp}). Section~\ref{s:Coda} summarizes these results and conclusions.

Adopting the notation of Refs.\cite{r1001,rHSS}, we list here a few basic definitions and results for completeness.
 $(2,2)$-extended supersymmetry on the $(1,1)$-dimensional worldsheet may be studied in $(1,1|2,2)$-superspace notation, wherein the supersymmetry charges $Q_\pm$ and $\Qb_\pm$, superderivatives $D_\pm,\Db_\pm$ and light-cone worldsheet derivatives $\vd_\pp,\vd_\mm$ satisfy:
\begin{subequations}
 \label{e:SuSy}
\begin{alignat}{3}
  \big\{\, Q_- \,,\, \Qb_- \,\big\}&=2i\vd_\mm, &\qquad
  \big\{\, Q_+ \,,\, \Qb_+ \,\big\}&=2i\vd_\pp,  \label{eSusyQ}\\
  \big\{\, D_- \,,\, \Db_- \,\big\}&=2i\vd_\mm, &\qquad
  \big\{\, D_+ \,,\, \Db_+ \,\big\}&=2i\vd_\pp,  \label{eSusyD}\\
  \vd_\mm&\Defl(\vd_\t{-}\vd_\s), &\qquad
  \vd_\pp&\Defl(\vd_\t{+}\vd_\s),
\end{alignat}
where $H{=}i\hbar\vd_\t$ and $p{=}-i\hbar\vd_\s$ are the worldsheet Hamiltonian and linear momentum, respectively. All other (anti)commutators among these operators vanish. In this notation, all operators\eq{e:SuSy} are eigen-operators of the Lorentz transformation and the number of ``$\pm$'' subscripts counts the additive eigenvalue in units of $\inv2\hbar$. So, the Lorentz-eigenvalue (``spin'') of $Q_\pm$ and $\Qb_\pm$ is $\pm\inv2\hbar$, of $\vd_\pp$ is $+\hbar$, and of $\vd_\mm$ is $-\hbar$.
\end{subequations}

As customary in superspace\cite{rSSSS1,r1001,rPW,rWB,rBK}, we use the super-derivatives $D_+,\Db_+,D_-,\Db_-$. With a nodding familiarity of {\em\/Adinkras\/}\cite{rA,r6-1,r6-3,r6-3.2} and similar diagrams\cite{rFre,rSIg0,rSIg}, we depict their algebra, up to $\vd_\pp,\vd_\mm$ factors, following Ref\cite{rHSS}:
\begin{equation}
 \vC{\begin{picture}(140,52)
      \put(-5,0){\includegraphics[width=150mm]{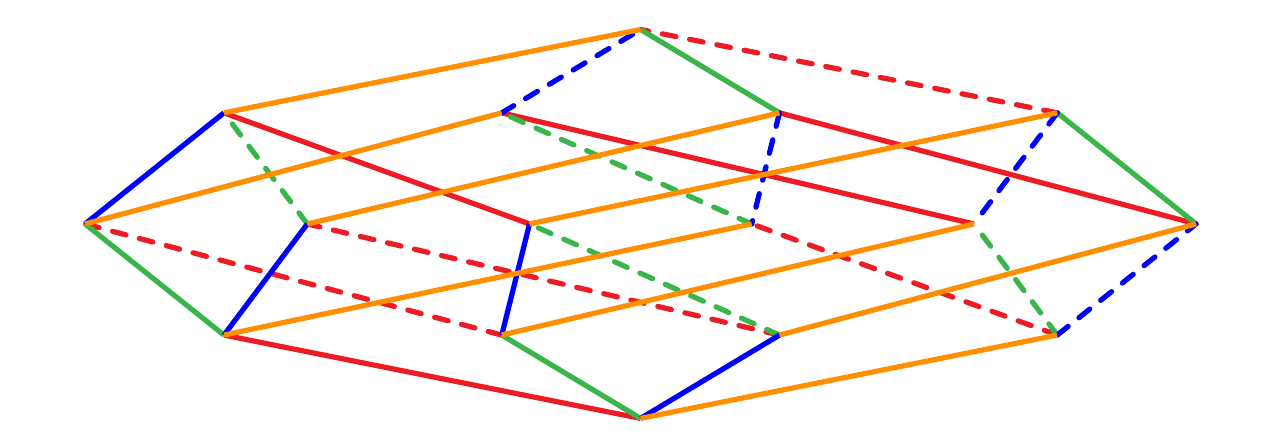}}
      \put(70,47.5){\cB{$\inv4[\Db_+,D_+][\Db_-,D_-]$}}
      \put(21,37.5){\cB{$\inv2[\Db_+,D_+]D_-$}}
      \put(54,37.5){\cB{$\inv2[\Db_+,D_+]\Db_-$}}
      \put(86.5,37.5){\cB{$\inv2[\Db_-,D_-]D_+$}}
      \put(119,37.5){\cB{$\inv2[\Db_-,D_-]\Db_+$}}
      \put(4.75,25){\cB{$\inv2[\Db_+,D_+]$}}
      \put(31,24.5){\cB{$D_+D_-$}}
      \put(57.5,24.5){\cB{$\Db_+D_-$}}
      \put(83.5,24.5){\cB{$\Db_-D_+$}}
      \put(110,24.5){\cB{$\Db_-\Db_+$}}
      \put(135.75,25){\cB{$\inv2[\Db_-,D_-]$}}
      \put(21,12){\cB{$D_+$}}
      \put(53.5,12){\cB{$\Db_+$}}
      \put(86.5,12){\cB{$D_-$}}
      \put(119,12){\cB{$\Db_-$}}
      \put(70,2){\cB{$\Ione$}}
     \end{picture}}
 \label{e:Ds}
\end{equation}
The Adinkras of Refs.\cite{rA,r6-1,r6-3,r6-3.2} depict representations of $N$-extended worldline supersymmetry, which is a subalgebra of\eq{e:SuSy} when $N\leq4$. Similarly here, following edges in\eqs{e:Ds}{e:Xs} {\em\/upward\/} corresponds to the supersymmetry action, where red/green/blue/orange edges correspond, respectively, to $\C1{D_+}$,- $\C2{\Db_+}$,- $\C3{D_-}$- and $\C4{\Db_-}$-action. Since $\Db_\pm=(D_\pm)^\dagger$, following an edges {\em\/downward\/} corresponds to the conjugate super-derivative action of the anti-color (green is anti-red, \etc), and incurs an additional $i\vd_\pp$ (for red/green edges) or $i\vd_\mm$ (for blue/orange edges) worldsheet derivative:
\begin{equation}
 \vC{\begin{picture}(30,25)(0,2)
       \put(0,0){\includegraphics[width=30mm]{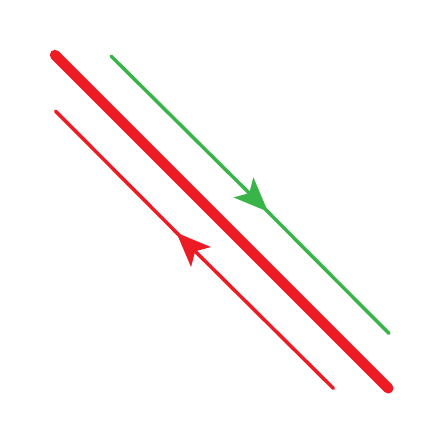}}
       \put(7,10){\C1{$D_+$}}
       \put(18,17){\C2{$\Db_+$}}
       \put(27,7){$i\vd_\pp$}
     \end{picture}}
  \qquad\textit{vs.}\quad
 \vC{\begin{picture}(30,25)(0,2)
       \put(0,0){\includegraphics[width=30mm]{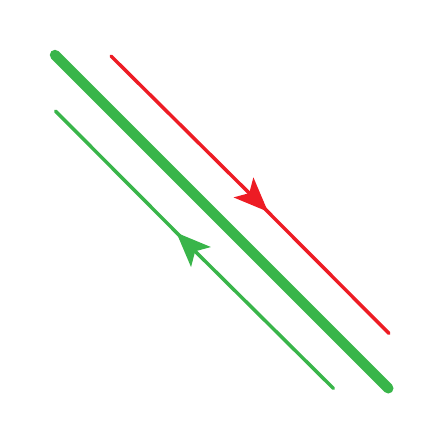}}
       \put(7,10){\C2{$\Db_+$}}
       \put(18,17){\C1{$D_+$}}
       \put(27,7){$i\vd_\pp$}
     \end{picture}}
  \qquad\textit{etc.}
  \label{e:UpDown}
\end{equation}
In\eq{e:Ds}, the node-labels clarify the action along any particular edge, \ie, between two nodes.

The components of any superfield|including superfield expressions and equations|are obtained by ``projecting'': applying the basis elements\eq{e:Ds} of the exterior algebra generated by $D_+.\Db_+,D_-,\Db_-$ and then setting the nilpotent superspace coordinates to zero. For an {\em\/intact\/}\ft{Herein, ``intact superfields'' are Salam-Strathdee complex superfields\cite{rSSSS1}, subject to neither constraint nor projection, all of which have a 4-cubical chromotopology\cite{r6-3} for worldsheet $(2,2)$-supersymmetry.} superfield $\BX$, we denote the so-obtained components:
\begin{equation}
 \vC{\begin{picture}(140,52)
      \put(-15,45){superfield $\BX$:}
      \put(-15,40){\small(intact)}
      \put(-5,0){\includegraphics[width=150mm]{Spindle.pdf}}
      \put(70,47.5){\cB{$\sX$}}
      \put(21,37.5){\cB{$\cX_+$}}
      \put(54,37.5){\cB{$\X_+$}}
      \put(86.5,37.5){\cB{$\cX_-$}}
      \put(119,37.5){\cB{$\X_-$}}
      \put(4.75,25){\cB{$X_\pp$}}
      \put(31,24.5){\cB{$X$}}
      \put(57.5,24.5){\cB{$\acute{X}$}}
      \put(83.5,24.5){\cB{$\grave{X}$}}
      \put(110,24.5){\cB{$\bar{X}$}}
      \put(135.75,25){\cB{$X_\mm$}}
      \put(21,12){\cB{$\c_+$}}
      \put(53.5,12){\cB{$\x_+$}}
      \put(86.5,12){\cB{$\c_-$}}
      \put(119,12){\cB{$\x_-$}}
      \put(70,2){\cB{$x$}}
     \end{picture}}
 \label{e:Xs}
\end{equation}
Comparing\eq{e:Xs} with\eq{e:Ds} shows that, for example,
\begin{equation}
 x\Defl\BX|,\quad
 \x_+\Defl\Db_+\BX|,\quad
 \bar{X}\Defl\Db_-\Db_+\BX|,\quad
 \cX_+\Defl\inv2[\Db_+,D_+]D_-\BX|,\quad\etc
 \label{e:fewXs}
\end{equation}
 These diagrams reveal the underlying 4-cubical (tesseract) {\em\/chromotopology\/}\cite{r6-3}\ft{For our purposes, the 4-cube (tesseract) is evident in the diagrams\eq{e:Ds} and\eq{e:Xs}, the edge-colors being correlated with the four dimensions of the 4-cube, and also the four operators, $D_+,\Db_+,D_-,\Db_-$.} of the supermultiplet represented by the superfield $\BX$ and encode all  supersymmetry relations. For example, using\eq{e:UpDown}, 
\begin{equation}
  \C2{\Qb_+}\,x=\x_+,\quad
  \C1{Q_+}\x_+=i(\vd_\pp x)-X_\pp,\quad
  \C2{\Qb_+}\,\x_+=0,\quad\etc,
 \label{e:QXs}
\end{equation}
where the minus sign in $\C1{Q_+}\x_+=\C1{Q_+}\C2{\Qb_+}x=-\inv2[\C2{\Qb_+},\C1{Q_+}]x+\inv2\{\C2{\Qb_+},\C1{Q_+}\}x$ is encoded by dashing the $\x_+$--$X_\pp$ edge, as in Ref.\cite{r6-1}.
 Depicting superfields in this manner reveals all the relations between its component fields|and indeed the whole structure of the superfield|in a correct but intuitive way that is not as easily conveyed by formulae.

\section{Strictly Constrained Off-Shell Superfields}
 \label{s:Strict}
There are two ways of constraining superfields: ({\small\bf1})~by defining them as satisfying a system of super-differential constraints, and ({\small\bf2})~by the inclusion of Lagrange multiplier terms in the Lagrangian.

Exploring systematically the former of these two methods, Ref.\cite{rHSS} defines, among many others, an ``almost unidexterous non-minimal superfield'' by imposing the second order super-differential condition:
\begin{equation}
 [\Db_-,D_-]\BA=0.
 \label{e:AL}
\end{equation}
Somewhat akin to\eq{e:L}, some of the components of this superfield vanish, while others satisfy worldsheet differential constraints:
\begin{subequations}
\begin{align}
 \{A_\mm;\cA_-,\A_-;\sA\}&=0,\label{e:AL0}\\
 \vd_\mm\,\{\a_-,\ra_-;A,\acute{A},\grave{A},\bar{A};\cA_+,\A_+\}&=0, \label{e:L8}\\
 \vd_\mm^{~2}\,\{a;\a_+,\ra_+;A_\pp\}&=0.\label{e:AL4}
\end{align}
\end{subequations}
with components defined akin to\eqs{e:Ds}{e:fewXs}, see\eq{e:[2]Ys}, and the indicated operators applied on each member of the indicated collection.
 Owing to the worldsheet differential constraints\eqs{e:L8}{e:AL4}, the $(1|4|5|2|0)$-component superfield defined to satisfy the super-differential constraint\eq{e:AL} then is not off-shell. In turn, the mapping that restricts an intact superfield to the one satisfying \Eq{e:AL} then is not a {\em\/strict homomorphism of off-shell supermultiplets\/}, as defined in Ref.\cite{r6-3.2}. The component fields of $\BA$ satisfying\eq{e:AL} are:
\begin{equation}
 \vC{\begin{picture}(135,52)
      \put(-5,0){\includegraphics[width=150mm]{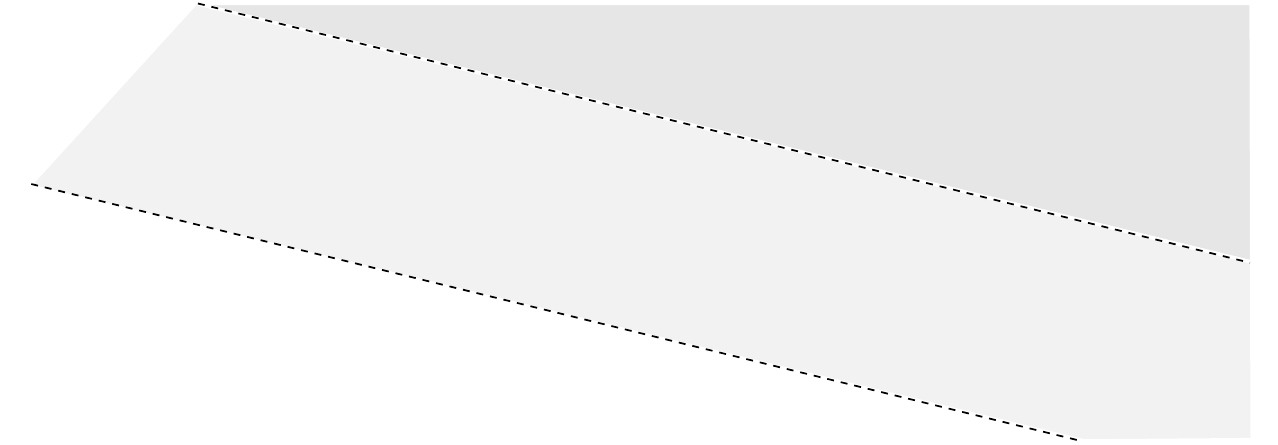}}
      \put(-5,0){\includegraphics[width=150mm]{Spindle.pdf}}
      \put(-18,48){$[\Db_-,D_-]\BA=0$:}
      \put(-18,43){\small(is not off-shell)}
       \put(0,5){\small\Eq{e:AL4}}
       \put(120,5){\small\Eq{e:L8}}
       \put(120,45){\small\Eq{e:AL0}}
      \put(70,47.5){\CB{\includegraphics[width=4mm]{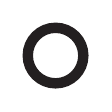}}}
      \put(21,37.5){\cB{$\2{\cA_+}$}}
      \put(54,37.5){\cB{$\2{\A_+}$}}
      \put(86.5,37.5){\CB{\includegraphics[width=4mm]{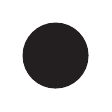}}}
      \put(119,37.5){\CB{\includegraphics[width=4mm]{Fermion.pdf}}}
      \put(4.75,25){\cB{$\2{\2{A_\pp}}$}}
      \put(31,24.5){\cB{$\2{A}$}}
      \put(57.5,24.5){\cB{$\2{\acute{A}}$}}
      \put(83.5,24.5){\cB{$\2{\grave{A}}$}}
      \put(110,24.5){\cB{$\2{\bar{A}}$}}
      \put(135.75,25){\CB{\includegraphics[width=4mm]{Boson.pdf}}}
      \put(21,12){\cB{$\2{\2{\a_+}}$}}
      \put(53.5,12){\cB{$\2{\2{\ra_+}}$}}
      \put(86.5,12){\cB{$\2{\a_-}$}}
      \put(119,12){\cB{$\2{\ra_-}$}}
      \put(70,2){\cB{$\2{\2a}$}}
     \end{picture}}
 \label{e:[2]Ys}
\end{equation}
where the unnamed open/closed nodes represent bosonic/fermionic expressions in terms of the other, named component fields. Simple underlining denotes a first order constraint\eq{e:L8}, whereas double underlining denotes the second-order constraint\eq{e:AL4}. There are no off-shell component fields in the $(1|4|5|2|0)$-component superfield\eq{e:[2]Ys}. The worldsheet differential constraints\eqs{e:L8}{e:AL4} obstruct using such a superfield in the path-integral formalism for quantum field theory.

By contrast, the non-minimal chiral (a.k.a.\ complex linear) superfield, defined below\eqs{e:CLM}{e:Ts}, is a proper, $(1|4|5|2|0)$-component off-shell superfield, as are also the non-minimal twisted chiral superfield, their Hermitian conjugates, and the more familiar chiral and twisted-chiral superfields.
\ping

The lefton superfields, defined by the pair of super-differential constraints in\eq{e:L}, are also not off-shell. Nevertheless, it turns out that precisely this superfield {\em\/on half-shell\/}\cite{rHP1} may be employed to eliminate the highlighted unnamed nodes indicated in\eq{e:[2]Ys}|without incurring any worldsheet differential constraints on the remaining component fields. That is, the (on half-shell) lefton superfield $\bL$ may be used to restrict an off-shell $(1|4|6|4|1)$-component intact superfield\eq{e:Xs} to a proper off-shell $(1|4|5|2|0)$-component superfield\eq{e:LYs}.
 This then also defines a strict morphism between these off-shell supermultiplets by the previously unexplored means of unidexterous Lagrange multipliers.

To see this, consider the Lagrangian term
\begin{equation}
 \sL_{\bL\BY}~\Defl~
 \inv4[\Db_+,D_+][\Db_-,D_-]\,\bL\,\BY|
 =L_\pp Y_\mm + \l_+\Y_- + \bl_+\cY_- + \ell\sY,
 \label{e:LAL}
\end{equation}
where
\begin{equation}
 \ell\Defl\bL|,\qquad
 \l_+\Defl D_+\bL|,\qquad
 \bl_+\Defl \Db_+\bL|,\qquad
 L_\pp\Defl \inv2[\Db_+,D_+]\bL|.
\end{equation}
are the components of the lefton superfield, $\bL$. With the standard assignment of mass-dimensions $[\bL]=0=[\BY]$, it follows that no kinetic term for $\bL$ is possible, so that all components of $\bL$ are {\em\/auxiliary\/}: their Euler-Lagrange equations are not differential\cite{rIPd,rHP1}. Since they occur linearly in\eq{e:LAL} and nowhere else, integrating out the components of $\bL$ imposes the simple constraints:
\begin{equation}
 Y_\mm~=~\Y_-~=~\cY_-~=~\sY~=~0,\qquad\textit{and nothing else!}
 \label{e:4}
\end{equation}
This defines a $\bL$-constrained version of $\BY$, which|unlike\eq{e:[2]Ys}|is a fully off-shell $(1|4|5|2|0)$-dimen\-si\-o\-nal representation of $(1,1|2,2)$-supersymmetry, spanned by the remaining, otherwise unconstrained component fields:
\begin{equation}
 \vC{\begin{picture}(140,52)
      \put(-15,45){Constr.\ by $\sL_{\bL\BY}:$}
      \put(-15,40){\small(is off-shell)}
      \put(-5,0){\includegraphics[width=150mm]{Spindle.pdf}}
      \put(70,47.5){\CB{\includegraphics[width=4mm]{Boson.pdf}}}
      \put(21,37.5){\cB{$\cY_+$}}
      \put(54,37.5){\cB{$\Y_+$}}
      \put(86.5,37.5){\CB{\includegraphics[width=4mm]{Fermion.pdf}}}
      \put(119,37.5){\CB{\includegraphics[width=4mm]{Fermion.pdf}}}
      \put(4.75,25){\cB{$Y_\pp$}}
      \put(31,24.5){\cB{$Y$}}
      \put(57.5,24.5){\cB{$\acute{Y}$}}
      \put(83.5,24.5){\cB{$\grave{Y}$}}
      \put(110,24.5){\cB{$\bar{Y}$}}
      \put(135.75,25){\CB{\includegraphics[width=4mm]{Boson.pdf}}}
      \put(21,12){\cB{$\h_+$}}
      \put(53.5,12){\cB{$\y_+$}}
      \put(86.5,12){\cB{$\h_-$}}
      \put(119,12){\cB{$\y_-$}}
      \put(70,2){\cB{$y$}}
     \end{picture}}
 \label{e:LYs}
\end{equation}
All named component fields are unaffected by the Lagrange multiplier constraining and remain off-shell. Thus, unidexterous constraining by means of the term\eq{e:LAL} turns out to define an off-shell superfield\eq{e:LYs}, which cannot be defined by means of a super-differential constraint such as\eq{e:AL}. The eliminated components, depicted by the highlighted and unnamed nodes, are connected by $\Qb_+,Q_+$ action into a quadrilateral. These features are exactly analogous to the non-minimal chiral superfield, defined below\eqs{e:CLM}{e:Ts}, its twisted variant and their Hermitian conjugates\cite{rHSS}.

The mapping $\BX\to\BY$ from an intact superfield\eq{e:Xs} to the one\eq{e:LYs} constrained by means of the Lagrangian term\eq{e:LAL} is then a strict homomorphism of off-shell supermultiplets, as defined in Ref.\cite{r6-3.2}. The mapping $\BX\to\BA$ specified by the second order super-constraint\eq{e:AL} is not.

\section{Unequal Off-Shell Bosonic and Fermionic Components}
 \label{s:Limp}
Consider a {\em\/non-minimal chiral\/} (a.k.a.\ complex-linear) superfield $\bQ$, defined by
\begin{subequations}
 \label{e:CLM}
\begin{gather}
 \Db_-\Db_+\bQ=0,\\[2mm]
 \vC{\begin{picture}(140,52)(-5,0)
      \put(-15,45){$\Db_-\Db_+\bQ=0:$}
      \put(-15,40){\small(is off-shell)}
      \put(-5,0){\includegraphics[width=150mm]{Spindle.pdf}}
      \put(70,47.5){\CB{\includegraphics[width=4mm]{Boson.pdf}}}
      \put(21,37.5){\cB{$\t_+$}}
      \put(54,37.5){\CB{\includegraphics[width=4mm]{Fermion.pdf}}}
      \put(86.5,37.5){\cB{$\t_-$}}
      \put(119,37.5){\CB{\includegraphics[width=4mm]{Fermion.pdf}}}
      \put(4.75,25){\cB{$T_\pp$}}
      \put(31,24.5){\cB{$T$}}
      \put(57.5,24.5){\cB{$\acute{T}$}}
      \put(83.5,24.5){\cB{$\grave{T}$}}
      \put(110,24.5){\CB{\includegraphics[width=4mm]{Boson.pdf}}}
      \put(135.75,25){\cB{$T_\mm$}}
      \put(21,12){\cB{$\q_+$}}
      \put(53.5,12){\cB{$\vq_+$}}
      \put(86.5,12){\cB{$\q_-$}}
      \put(119,12){\cB{$\vq_-$}}
      \put(70,2){\cB{$t$}}
     \end{picture}}
 \label{e:Ts}
\end{gather}
\end{subequations}
The component fields eliminated by the super-differential constraint (depicted as highlighted, unnamed nodes) are connected into a quadrilateral by $Q_+,Q_-$, \ie, $D_+,D_-$ action\cite{rFGH}.

Then, the inclusion of the Lagrangian term
\begin{equation}
 \sL_{\bL\bQ}\Defl\inv4[\Db_+,D_+][\Db_-,D_-]\,\bL\,\bQ|
 =L_\pp T_\mm + i\l_+(\vd_\mm\vq_+) + \bl_+\t_-
 \label{e:L-CLM}
\end{equation}
induces, through the equations of motion for $\bL$, only the constraints:
\begin{equation}
 T_\mm~=~(\vd_\mm\vq_+)~=~\t_-~=~0.
 \label{e:3}
\end{equation}
This turns the $(1|4|5|2)$-dimensional supermultiplet $\bQ$ into a $(1|4|4|1)$-dimensional one, where however one of the lowest-level fermions, $\vq_+$, has been rendered a left-mover, which is also the effect of its standard Dirac equation. This $\vq_+$ component field is therefore on-shell, whereas all other $(1|3|4|1)$-components remain off-shell. Therefore, this doubly constrained supermultiplet consists of 5 off-shell bosons and 4 off-shell fermions, coupled with one on-shell fermion:
\begin{equation}
 \vC{\begin{picture}(140,52)
      \put(-15,45){$\Db_-\Db_+\bQ=0$~\&~$\sL_{\bL\bQ}:$}
      \put(-15,40){\small(is not off-shell)}
      \put(-5,0){\includegraphics[width=150mm]{Spindle.pdf}}
      \put(70,47.5){\CB{\includegraphics[width=4mm]{Boson.pdf}}}
      \put(21,37.5){\cB{$\t_+$}}
      \put(54,37.5){\CB{\includegraphics[width=4mm]{Fermion.pdf}}}
      \put(86.5,37.5){\CB{\includegraphics[width=4mm]{Fermion.pdf}}}
      \put(119,37.5){\CB{\includegraphics[width=4mm]{Fermion.pdf}}}
      \put(4.75,25){\cB{$T_\pp$}}
      \put(31,24.5){\cB{$T$}}
      \put(57.5,24.5){\cB{$\acute{T}$}}
      \put(83.5,24.5){\cB{$\grave{T}$}}
      \put(110,24.5){\CB{\includegraphics[width=4mm]{Boson.pdf}}}
      \put(135.75,25){\CB{\includegraphics[width=4mm]{Boson.pdf}}}
      \put(21,12){\cB{$\q_+$}}
      \put(53.5,12){\CB{$\2{\vq_+}$}}
       \put(26,5){\small on shell}
       \put(21,1){\small (the only one)}
        \put(40,5){\vector(2,1){10}}
      \put(86.5,12){\cB{$\q_-$}}
      \put(119,12){\cB{$\vq_-$}}
      \put(70,2){\cB{$t$}}
     \end{picture}}
 \label{e:LTs}
\end{equation}
This time, the eliminated component fields (depicted by highlighted, unnamed nodes) are connected into two overlapping quadrilaterals. In addition, a single component field ($\vq_+$) is constrained on-shell by the combination of super-differential and Lagrange multiplier constraining; it occurs away from the eliminated quadrilaterals. The non-highlighted named component fields remain unconstrained and off-shell, but the single on-shell component $\vq_+$ prevents the entire superfield from being off-shell. A somewhat similar situation was noted in Ref.\cite{r6-4}, for which case however a coupling with another superfield was found to relax both of them off-shell. Whether such a completely off-shell extension of\eq{e:LTs} exists remains an open question.

\section{Conclusions}
 \label{s:Coda}
The preceding examples show that unidexterous ({\em\/on half-shell\/}\cite{rHP1}) Lagrange multiplier superfields can be used to define:
\begin{enumerate}\itemsep=-3pt\vspace{-4mm}
 \item proper off-shell superfields such as the new one%
\eq{e:LYs}, which cannot be defined by conventional super-differential constraining, as in\eq{e:[2]Ys};
 \item a strict homomorphism between the intact\eq{e:Xs} and the constrained off-shell superfield\eq{e:LYs};
 \item partially on-shell superfields, with unequal numbers of bosonic/fermionic off-shell component fields, such as\eq{e:Ts}.
\end{enumerate}\vspace{-3mm}
 Whereas this last feature does not seem to have an evident application, it is fairly novel and merits further study, perhaps \`a la Ref.\cite{r6-4}.
 On the other hand, the ability to define novel off-shell superfields such as\eq{e:LYs} is much more likely to have applications in constructions of novel worldsheet Lagrangians and models.
 In turn, the existence of the proper supersymmetry morphism~$(\ref{e:Xs}\to(\ref{e:LYs})$ may help a more rigorous understanding of off-shell supersymmetry representation theory.
\bigskip
\paragraph{Acknowledgment:}
 I am indebted to the generous support by the Department of Energy through the grant DE-FG02-94ER-40854, and wish to thank for the recurring hospitality and resources provided by
 the Physics Department of the University of Central Florida, Orlando, and
 the Physics Department of the Faculty of Natural Sciences of the University of Novi Sad, Serbia, where part of this work was completed.

\def\rasp{\leavevmode\raise.45ex\hbox{$\rhook$}}


\begin{thebibliography}{10}
 \expandafter\ifx\csname url\endcsname\relax
 \raggedright
 \def\url#1{\texttt{#1}}\fi
 \expandafter\ifx\csname urlprefix\endcsname\relax\def\urlprefix{ }\fi

\bibitem{rSSSS1}
A.~Salam, J.~Strathdee, {\em Supergauge transformations}, {\em Nucl. Phys.} B76
  (1974) 477--482.

\bibitem{r1001}
S.~J. Gates, Jr., M.~T. Grisaru, M.~Ro\v{c}ek, W.~Siegel, {\em Superspace},
  Benjamin/Cummings Pub. Co., Reading, MA, 1983.

\bibitem{rPW}
P.~West, {\em Introduction to supersymmetry and supergravity}, World Scientific
  Publishing Co. Inc., Teaneck, NJ, 1990.

\bibitem{rWB}
J.~Wess, J.~Bagger, {\em Supersymmetry and supergravity}, 2nd ed., Princeton
  Series in Physics, Princeton University Press, Princeton, NJ, 1992.

\bibitem{rBK}
I.~L. Buchbinder, S.~M. Kuzenko, {\em Ideas and methods of supersymmetry and
  supergravity}, Studies in High Energy Physics Cosmology and Gravitation, IOP
  Publishing Ltd., Bristol, 1998.

\bibitem{r6-1}
C.~F. Doran, M.~G. Faux, S.~J. Gates, Jr., T.~H\"{u}bsch, K.~M. Iga, G.~D.
  Landweber, {\em On graph-theoretic identifications of {A}dinkras,
  supersymmetry representations and superfields}, {\em Int. J. Mod. Phys.} A22
  (2007) 869--930.
\urlprefix\hbox{\url{http://arxiv.org/abs/math-ph/0512016}}

\bibitem{r6-3}
C.~F. Doran, M.~G. Faux, S.~J. Gates, Jr., T.~H\"{u}bsch, K.~M. Iga, G.~D.
  Landweber, R.~L. Miller, {\em Topology types of {A}dinkras and the
  corresponding representations of ${N}$-extended supersymmetry}.
\urlprefix\hbox{\url{http://arxiv.org/abs/0806.0050}}

\bibitem{r6-3.2}
C.~F. Doran, M.~G. Faux, S.~J. Gates, Jr., T.~H\"{u}bsch, K.~M. Iga, G.~D.
  Landweber, R.~L. Miller, {\em Adinkras for clifford algebras, and worldline
  supermultiplets}.
\urlprefix\hbox{\url{http://arxiv.org/abs/0811.3410}}

\bibitem{rFKS}
M.~Faux, D.~Kagan, D.~Spector, {\em Central charges and extra dimensions in
  supersymmetric quantum mechanics}.

\bibitem{rFS1}
M.~Faux, D.~Spector, {\em Duality and central charges in supersymmetric quantum
  mechanics}, {\em Phys. Rev. D (3)} 70(8) (2004) 085014, 5.

\bibitem{rFS2}
M.~Faux, D.~Spector, {\em A {BPS} interpretation of shape invariance}, {\em J.
  Phys.} A37 (2004) 10397--10407.

\bibitem{rKRT}
Z.~Kuznetsova, M.~Rojas, F.~Toppan, {\em Classification of irreps and
  invariants of the ${N}$-extended supersymmetric quantum mechanics}, {\em
  JHEP} 03 (2006) 098.
\urlprefix\hbox{\url{http://arxiv.org/abs/hep-th/0511274}}

\bibitem{rBKMO}
S.~Bellucci, S.~Krivonos, A.~Marrani, E.~Orazi, {\em `{R}oot' action for {N}=4
  supersymmetric mechanics theories}, {\em Phys. Rev. D} 73 (2006) 025011.

\bibitem{rBG}
S.~Bellucci, S.~J. Gates, Jr., E.~Orazi, {\em A journey through garden
  algebras}, {\em Lect. Notes Phys.} 698 (2006) 1--47.

\bibitem{rBKLS}
S.~Bellucci, S.~Krivonos, O.~Lechtenfeld, A.~Shcherbakov, {\em Superfield
  formulation of nonlinear {N}=4 supermultiplets}.

\bibitem{rHKLR}
N.~J. Hitchin, A.~Karlhede, U.~Lindstr{\"o}m, M.~Ro{\v{c}}ek, {\em
  Hyper-{K}{\"a}hler metrics and supersymmetry}, {\em Comm. Math. Phys.} 108(4)
  (1987) 535--589.

\bibitem{rHSS}
T.~H\"{u}bsch, {\em Haploid (2,2)-superfields in 2-dimensional space-time},
  {\em Nucl. Phys.} B555(3) (1999) 567--628.

\bibitem{rGSS}
R.~Q. Almukahhal, T.~H\"{u}bsch, {\em Gauging {Y}ang-{M}ills symmetries in
  1+1-dimensional space-time}, {\em Int. J. Mod. Phys. A} 16(29) (2001)
  4713--4768.

\bibitem{rHP1}
T.~H\"{u}bsch, I.~E. Petrov, {\em Worldsheet matter superfields on half-shell}.
\urlprefix\hbox{\url{http://arxiv.org/abs/0912.1038}}

\bibitem{rA}
M.~Faux, S.~J. Gates, Jr., {\em Adinkras: A graphical technology for
  supersymmetric representation theory}, {\em Phys. Rev. D (3)} 71 (2005)
  065002.
\urlprefix\hbox{\url{http://arxiv.org/abs/hep-th/0408004}}

\bibitem{rFre}
P.~Fr{\'e}, {\em Introduction to harmonic expansions on coset manifolds and in
  particular on coset manifolds with {K}illing spinors}, {\em in: Supersymmetry
  and supergravity \rasp 84 (Trieste, 1984)}, World Sci. Publishing, Singapore,
  1984, pp. 324--367.

\bibitem{rSIg0}
S.~Ichinose, {\em Graphical representation of supersymmetry}.
\urlprefix\hbox{\url{http://arXiv.org/abs/hep-th/0301166}}

\bibitem{rSIg}
S.~Ichinose, {\em Graphical representation of supersymmetry}.
\urlprefix\hbox{\url{http://arXiv.org/abs/hep-th/0603214}}

\bibitem{rIPd}
I.~E. Petrov, {\em On unidexterous matter and gauge superfields}, Ph.D. thesis,
  Howard University (2006).

\bibitem{rFGH}
M.~G. Faux, S.~J. Gates, Jr., T.~H\"{u}bsch, {\em Effective symmetries of the
  minimal supermultiplet of ${N} = 8$ extended worldline supersymmetry}, {\em
  J. Phys.} A42 (2009) 415206.
\urlprefix\hbox{\url{http://arxiv.org/abs/0904.4719}}

\bibitem{r6-4}
C.~F. Doran, M.~G. Faux, S.~J. Gates, Jr., T.~H\"{u}bsch, K.~M. Iga, G.~D.
  Landweber, {\em On the matter of ${N}=2$ matter}, {\em Phys. Lett. B} 659
  (2008) 441--446.
\urlprefix\hbox{\url{http://arxiv.org/abs/0710.5245}}

\end{thebibliography}
\end{document}